\documentclass[Journal]{vgtc}  

\onlineid{0}

\vgtcpapertype{Evaluation}

\title{Analyzing the Impact of Augmented Reality Head-Mounted Displays on Workers' Safety and Situational Awareness in Hazardous Industrial Settings}

\author{%
  \authororcid{Graciela Camacho-Fidalgo}{0009-0009-2978-9408}, 
  \authororcid{Blain Judkins}{0009-0006-0039-0063}, \authororcid{Kylee Friederichs}{0009-0000-4007-7481}, \authororcid{Lara Soberanis}{0009-0001-6862-6750}, \\ 
  \authororcid{Vicente Hernandez}{0009-0006-0226-4490}, \authororcid{Kevin McSweeney}{0000-0001-9135-0425}, \authororcid{Freddie Witherden}{0000-0003-2343-412X}, 
  \authororcid{Edgar Rojas-Mu\~noz}{0000-0001-6909-375X}
}

\abstract
{
Augmented Reality Head-Mounted Displays (AR-HMDs) have proven effective to assist workers. However, they may degrade their Safety and Situational Awareness (SSA), particularly in complex and hazardous industrial settings. This paper analyzes, objectively and subjectively, the effects of AR-HMDs' on workers' SSA in a simulated hazardous industrial environment. Our evaluation was comprised of sixty participants performing various tasks in a simulated cargo ship room while receiving remote guidance through one of three devices: two off-the-shelf AR-HMDs (Trimble XR10 with HoloLens 2, RealWear Navigator 520), and a smartphone (Google Pixel 6). Several sensors were installed throughout the room to obtain quantitative measures of the participants' safe execution of the tasks, such as the frequency in which they hit the objects in the room or stepped over simulated holes or oil spills. The results reported that the Trimble XR10 led to statistically highest head-knockers and knee-knocker incidents compared to the Navigator 520 and the Pixel 6. Furthermore, the Trimble XR10 also led to significantly higher difficulties to cross hatch doors, lower perceived safety, comfort, perceived performance, and usability. Overall, participants wearing AR-HMDs failed to perceive more hazards, meaning that safety-preserving capabilities must be developed for AR-HMDs before introducing them into industrial hazardous settings confidently.

}


\teaser{
  \centering
  \includegraphics[width=\linewidth, alt={A view of a city with buildings peeking out of the clouds.}]{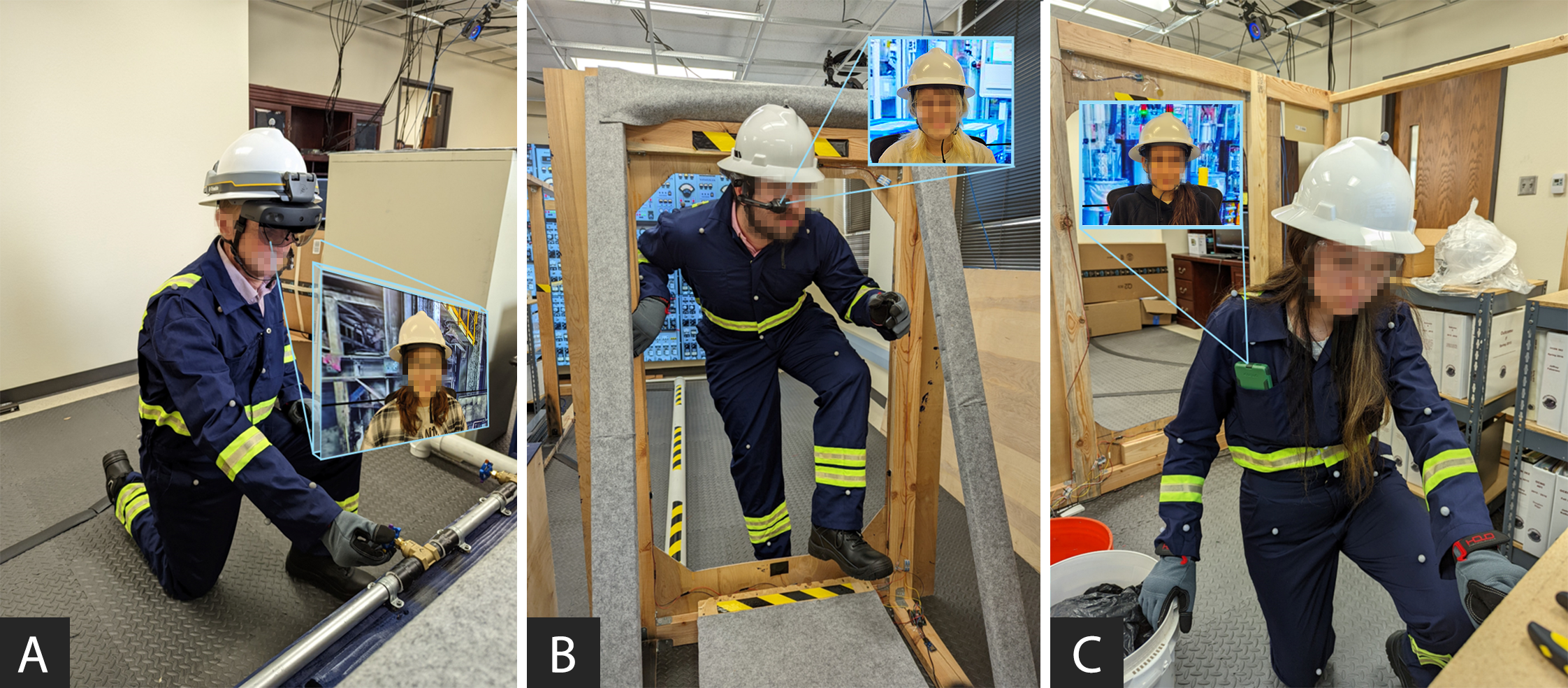}
  \caption{%
  	Participants performed various tasks within a simulated industrial hazardous environment while using either one of two off-the-shelf Augmented Reality Head-Mounted Displays (\textbf{A} - Trimble XR10 with HoloLens 2, \textbf{B} - RealWear Navigator 520), or a smartphone (\textbf{C} - Google Pixel 6). The user evaluation acquired quantitative measurements of the participants' safely execution of the tasks, along with qualitative insights of their perceived safety.
  }
  \label{fig:teaser}
}

\graphicspath{{figs/}{figures/}{pictures/}{images/}{./}} 

\usepackage{tabu}                      
\usepackage{booktabs}                  
\usepackage{lipsum}                    
\usepackage{mwe}                       
\usepackage{comment}

\usepackage{float}

\usepackage{color, colortbl}
\definecolor{LightGray}{rgb}{0.88,1,1}

\usepackage{mathptmx}                  

\begin{document}


\firstsection{Introduction}
\maketitle

Augmented Reality Head-Mounted Displays (AR-HMDs), when used within industrial settings, have demonstrated potential to reduce error rates, improve task execution, and boost productivity \cite{Hao2020SmartAugmentedRealityInstructionalSystemForMechanicalAssembly,DalleMuraCarbodyAssembly}, all while keeping the workers' hands unencumbered \cite{Wei2023HeadMountedDisplayAugmentedRealityInManufacturing}. Albeit effective to increase performance, studies have shown that AR-HMDs can also lead to a degradation of the workers' perception in complex environments \cite{Hietanen2020AR-BasedInteractionForHumanRobotCollaborative}. For instance, introducing AR imagery can distract workers from their immediate surroundings \cite{Sunwook2016AugmentedRealitySmartGlassesIndustryPerspective}, and negatively impact their visual information processing \cite{Drouot2022AugmentedRealityOnIndustrialAssemblyLineEffectivenessAndMentakWorkload}. As a result, cognitive demands increase and situational awareness is affected \cite{Kruijff2010PerceptualIssuesInAugmentedReality,Drouot2022AugmentedRealityOnIndustrialAssemblyLineEffectivenessAndMentakWorkload}. 

Addressing this disjunction between improved performance and degraded safety is of utmost importance to increase the acceptance rate of AR-HMDs within industrial settings. Although studies have examined AR-HMD's effects on safety and situation awareness (SSA), they have primarily focused on static workstations or subjective metrics with single hazards \cite{PinCognitive}. Therefore, there is a need for studies that objectively measure workers' SSA when using AR-HMDs within complex environments with multiple hazards, e.g., slippery platforms, tripping pipes, uneven surfaces \cite{Aromaa2020AwarenessOfTheRealWorldEnvironmentWhenUsingARHMD}.

The goal of this paper is to analyze whether current AR-HMD technology can be worn confidently in industrial hazardous environments without compromising the users' SSA. This goal is evaluated through a user study that quantifies the number of potential hazards users would have run into while wearing AR-HMDs within a simulated hazardous industrial environment. Sixty participants received remote guidance to complete various tasks involving navigation, decision-making, and object manipulation in a cargo ship room that included commonly existing hazards. The remote guidance was received through either of two off-the-shelf AR-HMDs solutions (Trimble XR10 with HoloLens 2, RealWear Navigator 520), or a hands-free smartphone (Google Pixel 6). Objective and subjective measurements of SSA were acquired through sensors installed across the room and questionnaires. Furthermore, our study seeks to address three main research questions:  

\begin{itemize}
    \item \textbf{RQ1}: Does the use of AR-HMD technology make users more prone to run into hazardous situations while completing industrial tasks?
    \item \textbf{RQ2}: Does the type of AR-HMD device have an effect in the users' ability to recognize hazards in an industrial setting?
    \item \textbf{RQ3}: Does the type of AR-HMD device have an effect in the users' self-reported feelings of safety and satisfaction? 
\end{itemize}

\section{Related Work}

Operations in industrial settings are primarily supported by paper-based practices and handheld devices (e.g., smartphones and tablets) \cite{BlauhutMobileDevices}. Albeit effective, the approaches can introduce safety risks to workers that are typically linked to the workspace encumbrance introduced by these approaches: they prevent the users from using both their hands and block their visual field \cite{DANIELSSONOperatorsperspective, SYBERFELDTShopfloor}. Therefore, there is a growing interest in alternatives to provide workers with guidance with an lower encumbrance footprint \cite{Wei2023HeadMountedDisplayAugmentedRealityInManufacturing}.

AR-HMDs have been explored as a successful alternatives to address such need: the devices offer similar capabilities to handheld devices without requiring operators to have their hands occupied \cite{Bottanni2019AugmentedRealityInManufacturingIndustry, Palmarini2018ASistematicReviewOfAugmentedRealityAplicationInMaintenance}. For instance, companies like BMW and Bosch have integrated AR technologies into their facilities to overlay digital schematics onto real-world machinery in fixed workstations \cite{Werrlich2017DemandAnalysis, Eswaran2023AugmentedRealityBasedGuidanceInProductAssemblyAndMaintenance}. Additionally, AR has been extensively used to connect remote experts with local workers in controlled environments, facilitating more effective real-time guidance and training. \cite{Aschenbrenner2019ComapringHumanFacctorsForAugmentedReality, Mourtzis2017CloudBasedAugmentedRealityRemoteMaintenance}.

The introduction of AR technologies into industrial settings, particularly AR-HMDs, should ensure that workers are not being subject to additional risks or distractions. As such, studies have attempted to measure awareness while using AR-HMDs in industrial settings. For instance, Aromaa et al.'s study reported that SSA was not significantly different between the Control and AR conditions \cite{Aromaa2020AwarenessOfTheRealWorldEnvironmentWhenUsingARHMD}. While their results were promising, their study has various limitations. For instance, they only acquired self-reported feelings of SSA without considering objective metrics. Additionally, their evaluation was conducted in a laboratory environment, where cognitive load is typically lower than in real-world scenarios. Furthermore, the task participants were asked to conduct (solving a puzzle) did not account for other factors such as physical demands and environmental stimuli. Howard et al.'s study expanded this by
conducting an evaluation in a simulated semi‐industrial facility \cite{Howard2023VisualInspectionWithAugmentedRealityHMD}. Their results pointed that workload demands in a visual inspection task were acceptable while using AR-HMDs. Nonetheless, the authors caution that these findings do not fully address the testing of all AR-HMD attributes in complex tasks or real industrial environments. Finally, Qin et al.'s research findings suggest that AR displays do not appear to significantly distract workers \cite{Qin2023MeasuringTheImpactOfARHMD}. However, their experimental conditions addressed only one type of hazard, i.e., tripping hazard, and tasks were performed in open areas with a lower presence of obstacles compared to the crowded and confined spaces present in complex industrial settings. 

The previous studies reveal that the use of AR-HMDs to provide support within industrial settings has various unanswered questions, particularly around the technology's impact in the workers' SSA. For instance validating the technology's usefulness have mostly been conducted in controlled laboratory settings \cite{Zhang}. The cognitive demands from such settings differ significantly from the ones of a dynamic industrial environment; the latter are often linked to high-risk conditions that could result in loss of life or severe injuries and are inherently more complex \cite{Silva2023MappingAccidentsAndUnsafeConditionsOfWorkersAutomativeSector,Song2016DynamicOccupationalRiskModelForOffshoreOperationsInHarhEnvironments,Jeelani2021RealTimeVisionBasedWorkerLocalization&HazardDetection}. Furthermore, most studies have been focused on evaluating AR techniques through performance metrics such as error rate and completion time \cite{Egger2020AugmentedRealityinSupportofIntelligentManufacturing, Bosch2017TheEffectsOfProyectedVersusDisplayInstructionsOnProductivity,Bottanni2019AugmentedRealityInManufacturingIndustry}, and often ignore the impact of AR on the workers' SSA \cite{Gerdenitsch2022AugmentedRealityAssitedAssembly}. The studies that do have provided SSA measures have done not mostly through subjective metrics that may be influenced by factors such as the technology's novelty of the technology \cite{Hussain2024ExploringConsttructionWorkersAttentionAndAwareness, Namgyun2021PredictingWorkersInattentivenessToStruckHazards, Qin2022MeasuringTheImpactOfInformationDisplayMethodsOnARHMD}. Our work addresses these questions and provides pointers for an effective integration of AR technology within industrial hazardous settings.

\section{Evaluation of AR-HMDs within industrial hazardous environments}

Our work proposes an user study to analyze whether current AR-HMD technology adequately ensures safety within industrial hazardous environments without compromising the situational awareness of users. Participants were recruited to perform various inspection tasks in a simulated hazardous environment while receiving remote expert guidance via Microsoft Teams on one of three devices: two off-the-shelf AR-HMDs (Trimble XR10 and Navigator 520), and a traditional smartphone approach (Google Pixel 6) (see Figure \ref{fig:teaser}). Quantitative and qualitative data on their SSA and satisfaction were collected using multiple sensors placed throughout the room, as well as pre and post-test questionnaires. 

\subsection {Simulating an Industrial Maritime Environment}

We selected the maritime industry as our testbed for an industrial hazardous setting. Specifically, our setup remodeled a 32.5 ft x 15.2 ft room to recreate a steering gear room from a cargo vessel. The area designated for the test tasks was covered with exercise mat flooring and maintained an average luminance condition of (750-830 lux). A constant background noise obtained from a real ship facility (70-80dB) was played using a speaker system installed in the ceiling (one JBL Commercial 70v Amp + four 6" Ceiling Speakers) to replicate the typical noise found in industrial settings. Furthermore, we conducted a literature review and received expert feedback from active marine professionals to design the environment, including elements that characterize complex industrial settings. These elements include confined spaces, interaction with complex equipment, navigation through crowded spaces, changes in surface elevation, and dynamic stimuli.

\begin{figure}
    \centering
    \includegraphics[width=1\linewidth]{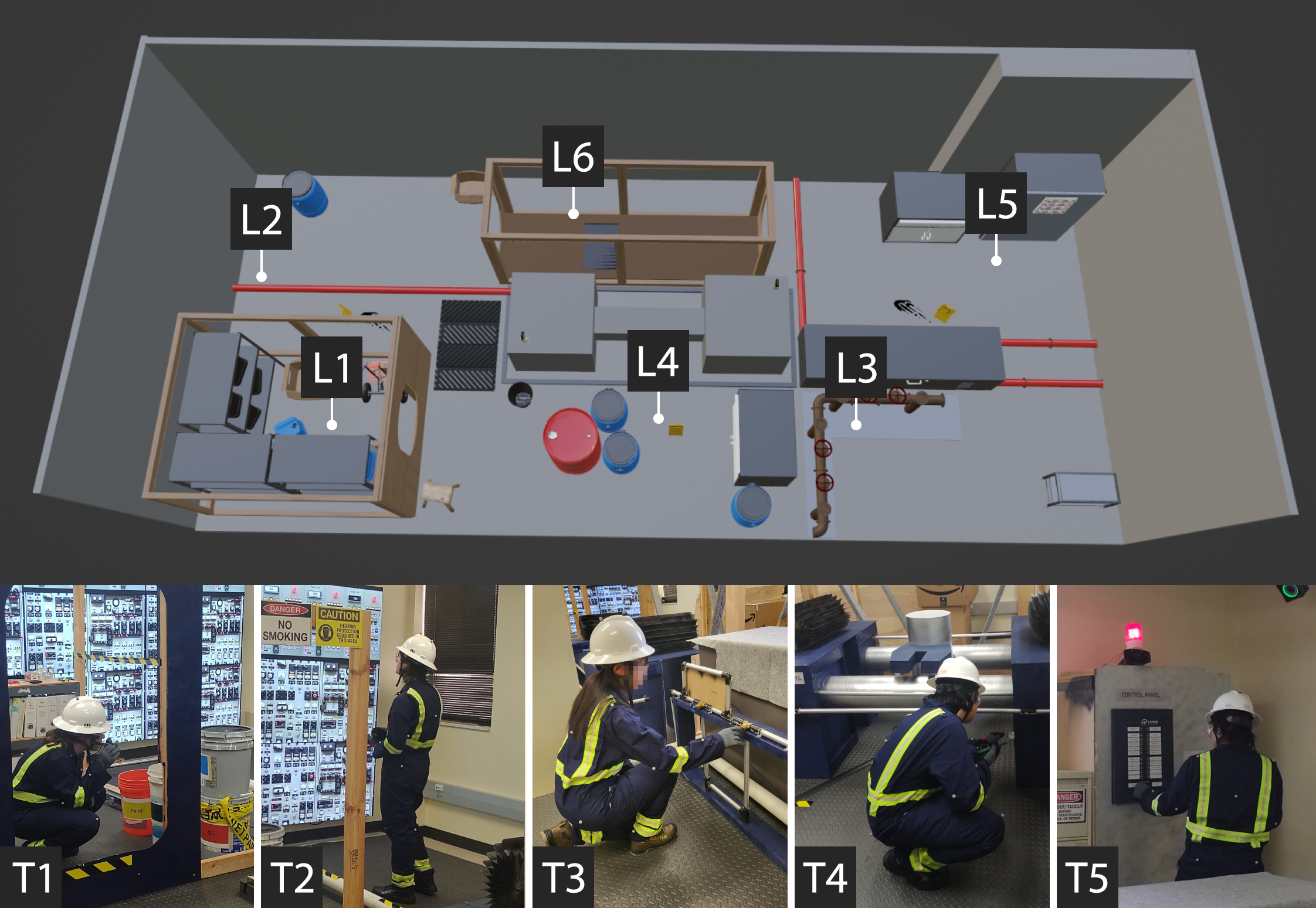}
    \caption{Tasks (T1-T5) and their associated locations (L1-L6).}
    \label{fig:tasks}
\end{figure}

The resulting physical building environment consists of six defined locations that participants traverse. As shown in Figure \ref{fig:tasks}, the locations are L1) Storage Room, L2) Panel Verification, L3) Emergency Pump System, L4) Steering Gear System, L5) Alarm Control Panel, and L6) Catwalk Platform. The Storage Room is held in a wooden frame structure (7.3 ft x 7.9 ft x 5.7 ft), with an entrance hatch door (31" wide by 51" high, with an elevation 10" off the ground). Inside, various inventory items are placed, such as binders, tools, and container buckets. The Steering Gear System, constructed from wood, features visual elements such as pressure gauges and pipes, resembling real equipment.The Panel Verification area utilizes six TV screen panels to display informational control settings in an immersive manner. The Catwalk Platform is an elevated wooden surface (3.2 ft x 6.5 ft x 5.7 ft, with an elevation of 10" of the ground) with hatch doors providing access to the other side of the room. The Alarm Control Panel consists of a board with a smart plug button (Kasa Smart Plug HS103P2) connected to a sound and light alarm (Jiawanshun's Cathy-046, 110V, 130dB), allowing the alarm to be activated or deactivated remotely or on-site. Lastly, the Emergency Pump System is constructed from 3/4" PVC pipes, turnable valves, and three installed status LCD lights. The spokes of the valves are painted in different colors for task-specific purposes.

\subsection {Designing Simulated Hazards}

\begin{figure} [ht]
    \centering
    \includegraphics[width=1\linewidth]{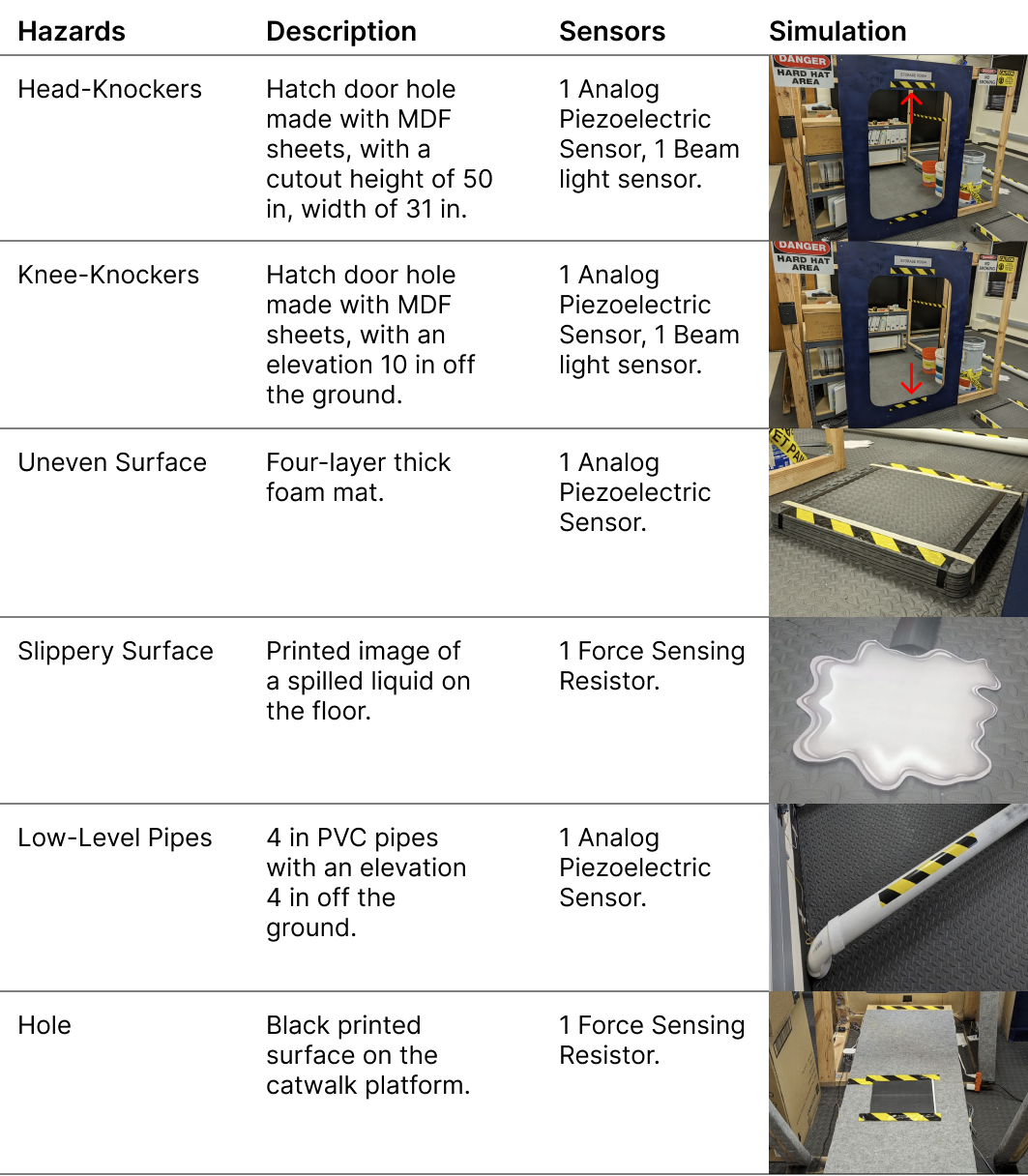}
    \caption{Description of hazards present in the room and their physical simulation, including sensors.}
    \label{fig:hazards}
\end{figure}

According to the Occupational Safety and Health Administration (OSHA), slip, trip, and fall accidents are the most common causes of injuries among maritime workers and in other industries \cite{OshaSlipsTripsFalls}. Hence, we compiled a list of different hazards related to the risk of slip, trip, and fall accidents among maritime operations, and selected the six most common hazards to be incorporated into our environment. A total of 14 simulated hazards were evaluated across these six categories: 3 Head Knockers, 3 Knee Knockers, 1 Uneven Surface, 2 Slippery Surfaces, 4 Low-level pipes, and 1 Hole. The selection criteria included the ability to safely replicate or simulate the hazards, prevalence in other complex environments, and the frequency of accidents mentioned in the literature. Figure \ref{fig:hazards} illustrates and describes the different types of hazards simulated in our industrial environment. In addition to these hazards, common environmental elements (e.g., boxes, ladders, drums, and buckets) were included to contextualize the setting and obstruct pathways. 

\subsection {Measuring of Hazard Perception}

To objectively measure the participants' hazard perception while performing tasks, 18 sensors were installed over the hazardous elements inside the room. These included Beam Light Sensors, Force Sensing Resistors, and Analog Piezoelectric Sensors. The quantity and type of sensors used to detect occurrences of hazardous situations (i.e., when participants come into contact with a hazard) for each hazard element are illustrated in Figure \ref{fig:hazards}. Each group of sensors per hazard element was controlled by an Arduino Nano ESP32 \cite{Arduino} connected to an SD card module. The generated data (i.e., Hazard, Location, ArduinoID, SensorID, SensorData, Date, Time) was stored locally on an SD card as a text file, with a delay frequency of 50 milliseconds between line recordings. Additionally, all the Arduinos were connected to a smart plug (Kasa Smart Plug HS103P2) to remotely control the start and end timestamps of each participant test using a Python code with the python-kasa library \cite{KostonKasa}.

In addition to the sensors, a motion capture system was installed in the room to evaluate participants' movements throughout the experiment. The motion capture area was designed using 12 Prime\textsuperscript{x} 22 OptiTrack Motion Tracking cameras \cite{OptiTrack} positioned along the room’s perimeter, capturing data at 120 frames per second. Since the objective of the present work was to evaluate hazard perception rather than directly assess the participants' physical performance, the data obtained from the motion capture system was primarily used to calculate completion times and to address any sensor failures, with additional applications planned for future analysis.

\subsection{Performing Industrial Tasks}

Participants were required to take on the role of a novice marine surveyor to complete five tasks involving navigation, exploration, decision-making, and object manipulation, all while being mindful of common hazards found in such environments. The task design aimed to simulate a complex scenario with a high level of cognitive load, mirroring the dynamism and complexity of real hazardous environments. The task design was also validated by active marine personnel to accurately reflect common maritime tasks. 

Throughout the experiment, participants used the Microsoft Teams application to communicate with a remote agent (i.e., a member of the research team, located in an adjacent room) to receive guidance and access files required to complete the tasks. The remote agent assumed the role of an expert marine surveyor to guide and assist the participants during their training sessions as novice marine surveyors. Participants could consult the agent for clarifications on task execution and were required to provide status updates for each task performed during the test. This scenario simulated a remote version of the current in situ training, where an expert operator accompanies trainees during routine tasks to provide support and assistance. The objective of this approach was to gather insights into aspects of remote assistance using different AR-HMD off-the-shelf devices with the Microsoft Teams app, which is widely used for remote assistance in industrial settings. 

The five simulated tasks designed are illustrated and located in Figure \ref{fig:tasks}. Participants had to complete these five tasks and return to their original location to finish the experiment. The tasks were: \textbf{1) Check inventory items}, \textbf{2) Report values from a control panel}, \textbf{3) Turn fire valves}, \textbf{4) Check pressure gauges}, and \textbf{5) Stop an alarm}. Additionally, participants were required to kneel, squat, and look at elevated areas on several occasions during the tasks, and they were required to wear Personal Protective Equipment to comply with industry safety standards (PPE, e.g., safety coveralls, safety glasses, disposable earplugs, steel-toe boots, hard hat, and safety gloves).

The \textbf{Check Inventory Items} task (Figure \ref{fig:tasks}, T1) required participants to access the Storage Room, locate various binders, tools, and buckets, and confirm their presence with the remote agent. The remote operator sent three images via Microsoft Teams chat, each listing the objects participants needed to find. Participants opened these files and reported any missing items to the remote agent. Next, the \textbf{Report Values from a Control Panel} task (Figure \ref{fig:tasks}, T2) required participants to navigate to the Panel Verification area and report the blue values displayed on the immersive TV panels to the remote agent. Then, in the \textbf{Turn Fire Valves} task (Figure \ref{fig:tasks}, T3) participants proceeded to the Emergency Pump System area, where they encountered three fire valves. They accessed another image sent by the remote agent via Microsoft Teams chat. The image depicted six valve configurations that participants needed to replicate by adjusting the valve spokes to match the orientations.

An emergency was simulated by setting an alarm off in the room. The alarm was set to go off three minutes after the experiment had begun via a Python script using the python-kasa library \cite{KostonKasa}. Participants had to stop their current task and immediately switch to the \textbf{Check pressure gauges} task (Figure \ref{fig:tasks}, T4). To complete this task, participants had to navigate back to the Steering Gear System area and verify that the readings from two pressure gauges located within the system were zero. Upon confirming this, participants proceeded to the \textbf{Stop an alarm} task (Figure ~\ref{fig:tasks}, T5). In this task, participants had to navigate to the Alarm Control Panel area and push a button to stop the alarm. After completing these two tasks, participants had to resume their previous task after the alarm.

\subsection{Experimental Conditions}

The experiment evaluated the objective and subjective differences in completing marine simulated surveying tasks using three devices: the Trimble XR10, the Navigator 520, and the Pixel 6. The Trimble XR10 and the Navigator 520 were selected based on a benchmark performed over current off-the-shelf AR-HMDs. The team inspected literature and industry recommendations \cite{Wei2023HeadMountedDisplayAugmentedRealityInManufacturing, GuoARMEASUREMNTES} to compare devices based on two main features: 1) ruggedness \& ability to handle outdoor environments, and 2) virtual imagery display capabilities. A two-dimensional coordinate system was created to map this feature space, and several off-the-shelf AR wearable devices were mapped into it. The devices were also expected to be untethered, intrinsically safe, conform to PPE, and should allow the user to be hands-free while using them. Furthermore, no new software or technology was expected to be developed for this experiment as the goal was to evaluate these current off-the-shelf AR-HMDs and their capabilities through common software suites such as Microsoft Teams. The Pixel 6 condition was selected based on its resemblance with current approaches to perform remotely-guided tasks. Therefore, the selected devices aim to represent different levels of AR capabilities among off-the-shelf AR wearable device options (i.e., monocular display, handled, stereo see-through display).

\begin{figure} 
    \centering
    \includegraphics[width=1\linewidth]{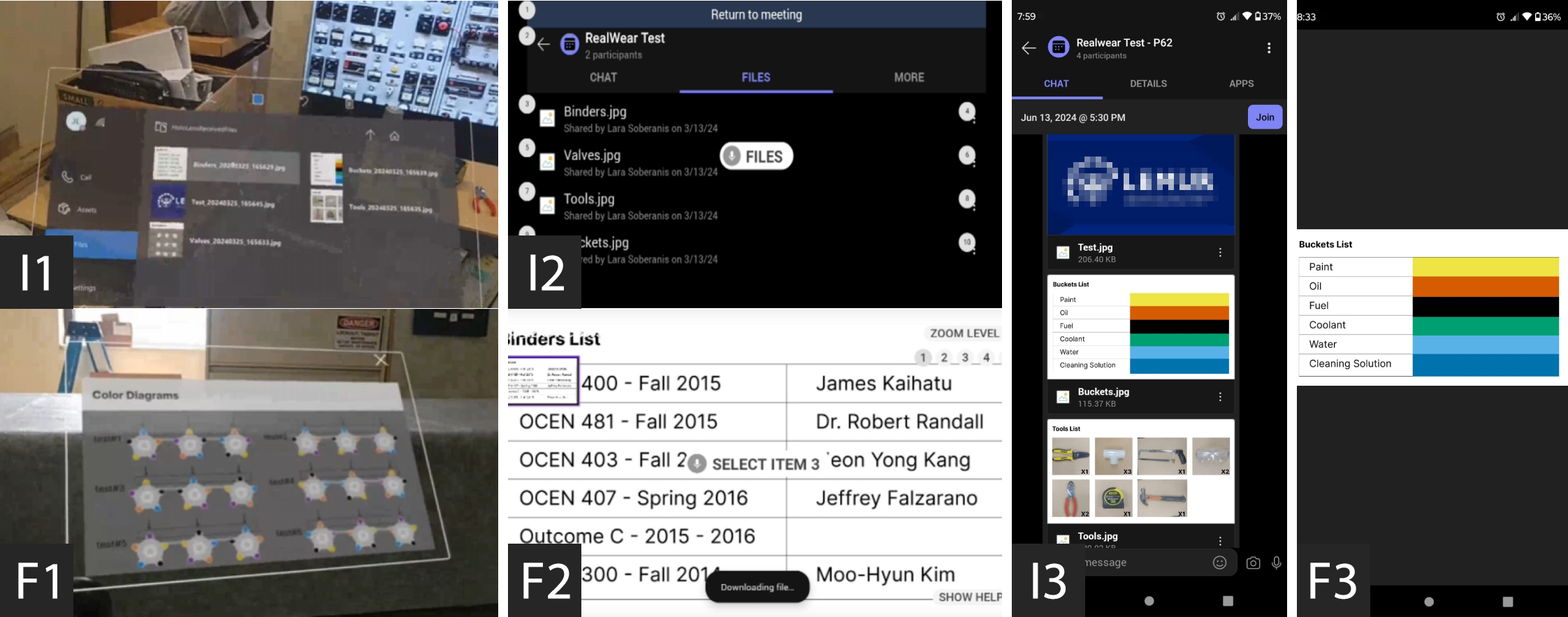}
    \caption{Microsoft Teams \textbf{interface} to access the files (I1-I3) and \textbf{first-person} view visualization of the files (F1-F3).}
    \label{fig:interfaces}
\end{figure}

The method of interacting with the Microsoft Teams interface to communicate with the remote agent and access the files for the tasks differed between conditions. For the Trimble XR10, participants relied on 3D virtual panels superimposed over their workspace. These panels could follow them or be pinned to the world space. To find and select the files, participants had to interact with the panels via air-tapping (Figure \ref{fig:interfaces}, I1 \& F1). Conversely, participants using the Navigator 520 visualized the interface and the files on the device's monocular display, positioned directly under their right eye. They used voice commands to navigate through the call and access the files (Figure \ref{fig:interfaces}, I2 \& F2). Finally, participants in the Pixel 6 condition used touch interactions to interact with the call interface and navigate through the files (Figure \ref{fig:interfaces}, I3 \& F3).

\subsection {Experiment Logistics}

Each participant was received at a common room prior to the experiment. A member of the research team provided an overview of the study objectives and invited participants to sign an informed consent form before proceeding. They proceeded to fill out one pre-experiment questionnaire. Afterwards, participants underwent a 30-minute training session (to be explained in the next subsection). Upon completing the pre-experiment training, participants entered the test room and completed the five aforementioned tasks. After completing the tasks, participants returned to the common room to complete various post-experiment questionnaires. The experiment was approximately 50 minutes long. A research team member was present at the perimeter of the simulated room as a precautionary measure to assist participants and prevent actual accidents. 

\subsection {Pre-Experiment Training}
 
To ensure safety and familiarity with the environment, participants underwent three custom-made training sessions before the experiment: \textbf{safety practices}, \textbf{personal protective equipment}, and \textbf{device usage}. These training sessions gave participants the fundamental knowledge needed to act as novice marine surveyors in the simulated hazardous environment. During the \textbf{safety practices} training, participants received an introduction to the marine surveyors' routine tasks, instruction on the circumstances of a hazardous workplace; the accident-prone situations they may encounter, and the best practices to avoid such incidents (e.g., Pay close attention to obstacles, Walk, don’t run, Don't drag or slide your feet along the ground, among others). Afterwards, participants received a \textbf{PPE} training to familiarize themselves with its usage. They were introduced to the PPE to be used and were requested to wear it correctly throughout the training session to enhance their familiarity with it. Finally, participants underwent a \textbf{device usage} training specific to their assigned condition. This training focused on interacting with the devices and accessing files for the tasks through the devices' Microsoft Teams application. Different interaction methods. These included hand gestures for the Trimble XR10 (pointing, grabbing, or swiping), voice commands for the Navigator 520 ('navigate back,' 'zoom level 2' 'select item 3', and 'freeze window.'), and touchscreen interactions for the Pixel 6 (tapping, swiping, and pinching).

\subsection{Comprehensive SSA Evaluation Metrics}

Each sensor installed in the room was organized into sensor groups, as illustrated in Figure \ref{fig:hazards}. The total number of activations for each sensor and group was recorded for every participant. For instance, if a participant kicked a low-level pipe, the corresponding sensor group would register an activation event. Two types of data pre-processing routines were incorporated to accurately identify sensor activations. First, a \textit{base activation threshold} was established to distinguish real activations from false positives and noise. To do this, multiple tests were conducted, deliberately triggering each sensor several times by tripping or hitting each hazard. This generated three correct activation values, which were then averaged. The base activation threshold was set to ignore events whose magnitudes were lower than a third of the averaged activation values. Subsequently, a \textit{time threshold} was established to prevent multiple counts of the same event, such as vibrations produced by a kick resulting in several activations within a short period. An empirical threshold of 3 seconds was applied, during which additional activations beyond the first were ignored. 

In addition to the objective measures, participants completed one pre-experiment questionnaire and several post-experiment questionnaires. The pre-experiment questionnaire (Demographics) recorded the participants' demographics, such as age range, familiarity with the use of PPE, and familiarity with AR technology, among others. The post-experiment questionnaires, two standard questionnaires were used to evaluate usability (System Usability Scale; SUS \cite{slater2000virtual,usoh2000using}), and perceived workload (Task Load Index; TLX \cite{hart1988development}), along with three custom-made questionnaires to evaluate the participants' self-reported feelings of safety (User Safety Perception), their ability to identify and recall hazards in the room (Hazard Identification) and the devices' ease-of-use (User Device Perception). 

\subsection{Statistical Analysis}

Our statistical analysis treated the devices as independent variables and all the aforementioned metrics as dependent variables. The null hypothesis for all comparisons was that all devices yield equal results across all metrics. The normality assumption of the data was assessed using the Shapiro-Wilk test \cite{shapiro1965analysis}. Afterward, Levene’s test was employed to evaluate the equality of variances  \cite{levene1960robust}. Finally, a Tukey-Kramer test was run to compare the conditions \cite{kramer1956extension,tukey1949comparing}.

\section{Results \& Discussion}

The experimental procedures were reviewed and approved by the Institutional Review Board (\textit{IRB2023-1223}), and written informed consent was obtained from all participants prior to the experiment. A total of 60 participants (22 females, 37 males, 1 non-binary, age = 28.30 ± 10.17 years) were recruited and randomly assigned to one of three experimental conditions. Fourteen people reported familiarity with ship facilities, 13 reported some awareness of the safety protocols and procedures typically followed in ship facilities, 2 reported experience with marine surveyor tasks on vessels, 51 reported some experience with Personal Protective Equipment (PPE), 23 reported having intermediate to expert level experience with Augmented Reality (AR) technology, and 53 reported having average to excellent multitasking ability. 

\subsection{Sensor Activations}

For the quantitative evaluation of safety perception, sensor activations were categorized according to the type of hazard: Head Knockers, Knee Knockers, Uneven Surfaces, Slippery Surfaces, Holes, and Low-Level Pipes. Additionally, the activation data from the Beam Light sensors, which detected when participants crossed the hatch doors (used as a reference for the Head Knockers and Knee Knockers groups), were considered as an additional group for the Tukey-Kramer test, as they did not directly detect hazardous situations but provided indicators of SSA.

The results indicate that participants using the Trimble XR10 hit their heads (head-knocker hazard) significantly more often in the storage room (1.20 ± 1.79) compared to those using the Navigator 520 (0.25 ± 0.44; p = 0.02) and the Pixel 6 (0 ± 0; p < 0.01). Similarly, Trimble XR10 users encountered hit their feet and knees (knee-knocker hazard) significantly more often in the storage room (0.40 ± 0.68) compared to Navigator 520 users (0.05 ± 0.22; p = 0.04) and Pixel 6 users (0.05 ± 0.22; p = 0.04). Additionally, Trimble XR10 participants also hit their head more frequently on the Platform right side (7.20 ± 5.98) than those using the Navigator 520 (3.60 ± 2.64; p = 0.02) and the Pixel 6 (3.05 ± 1.90; p < 0.01). 

\begin{figure*}[tb]
  \centering 
\includegraphics[width=0.85\linewidth]{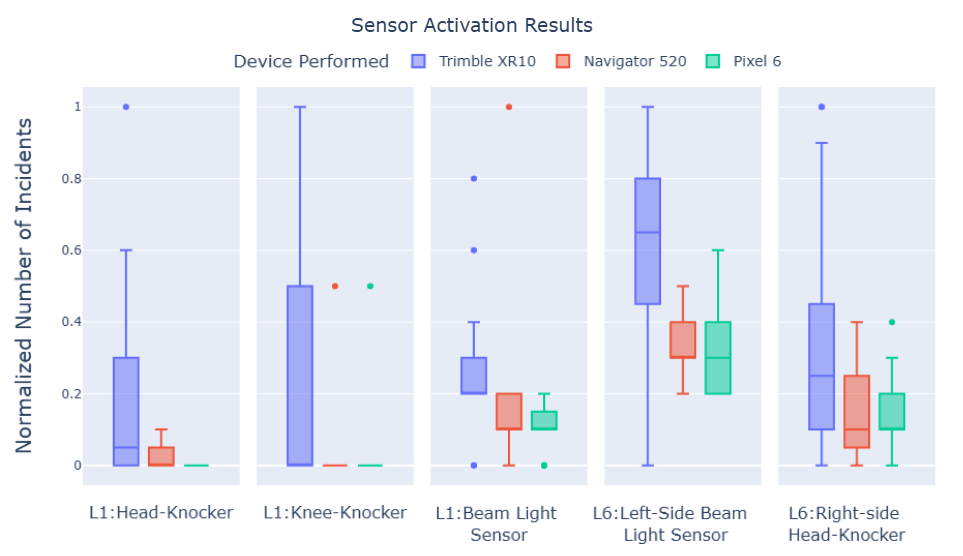}
\captionof{figure}{Sensor normalized number of incidents that reported statistically significant differences between the conditions. Overall, the Trimble XR10 led to significantly more sensor activations.}
\label{fig:sensorActivations}
\end{figure*}

For the Beam Light sensor group, Trimble XR10 participants crossed the storage room hatch door significantly more often (50.40 ± 34.69) compared to those using the Navigator 520 (32.70 ± 36.69; p = 0.02) and the Pixel 6 (20.70 ± 10.36; p < 0.01). Similarly, the Trimble XR10 condition reported significantly more activations of the platform left side hatch door (36.75 ± 15.43) compared to Navigator 520 users (19.45 ± 5.34; p < 0.01) and Pixel 6 users (19.50 ± 7.10; p < 0.01). Figure \ref{fig:sensorActivations} depicts the data for Head-Knockers, Knee-Knockers, and the Light Beam sensor group. No other statistically significant differences were observed between conditions.

The results of the objective measurements indicate that the use of AR-HMDs led to a significant reduction in the ability to prevent head and knee incidents while performing industrial tasks (\textbf{RQ1}). Previous research found no significant differences in hazardous encounters between devices when simulating tripping hazards \cite{Qin2023MeasuringTheImpactOfARHMD}. Our study reached a similar conclusion, showing no significant differences in tripping hazards. However, we assessed other types of simulated hazards and found meaningful differences in Knee-Knocker and Head-Knocker incidents. This discrepancy may suggest that these specific hazards are influenced by additional factors not considered in prior research. For instance, the weight of the device, as discussed by \cite{Constantino2021NewAnEmergingHazardsForHealthAndSafetyWithinDigitalizedManufacturing}, can affect user perception and navigation due to the discomfort it causes, which increases cognitive demands. This heightened cognitive load becomes especially significant in scenarios where users need to crouch and gauge their head position relative to hatch doors or elevated objects, thereby affecting their ability to avoid Head-Knockers and Knee-Knocker hazardous situations. 

Additionally, the Beam Light Sensor showed significantly highest activations in the Trimble XR10, which means they crossed hatch doors more frequently and with higher difficulty than those in other conditions. As observed by \cite{Howard2023VisualInspectionWithAugmentedRealityHMD}, Trimble XR10 users occasionally forgot to unpin or pin the call interface when moving between areas. This resulted in multiple crossings of the hatch doors to recover the interface, leading to increased activations of the Beam Light Sensors. Additionally, the complexity of the environment and the presence of acute stressors (e.g., sound and light alarms) may impair memory, causing participants to forget the required tasks and consequently move around the room more frequently. While this oversight may not directly impact user safety in short-term evaluations, it could pose risks in long-term use cases.

Our findings emphasize the need to consider device weight and hazard types in future studies to better understand their impact on hazard avoidance. Additionally, accounting for varying hazard risks is crucial for effective hazard mitigation strategies. Moreover, User Interface and Experience (UI/UX) designers should consider implementing better affordances to help users maintain awareness of panel pinning status (e.g., include larger icon statuses, incorporating a distinctive color to draw attention, or notifying users after they move a certain distance from a panel) \cite{Antonellaaffordances}. Furthermore, as a future direction, exploring systems capable of dynamically determining the optimal interface placement status based on contextual awareness of the environment and the user's cognitive load could be valuable. These adaptive designs are becoming increasingly relevant, with potential approaches including human cognitive behavior analysis and Artificial Intelligence \cite{KringsAdaptive}.

\subsection{User Device Perception - UDP}

A Likert scale questionnaire (1—Strongly Agree to 5—Strongly Disagree) was used to evaluate the subjective user perception of the devices, based on aspects like device efficiency, performance, well-being, comfort, productivity and learning promotion. Table~\ref{tab:SubjectiveSummaries} presents the summarized results for each UDP question.

Regarding efficiency (Question 2), participants using the Trimble XR10 felt significantly hindered in completing tasks quickly (0.60 ± 0.27) compared to the Navigator 520 (0.43 ± 0.18; p-value = 0.04) and the Pixel 6 (0.36 ± 0.18; p-value < 0.01). Similarly, in terms of productivity (Question 3), Trimble XR10 users reported less enhancement in productivity (0.58 ± 0.23) compared to the Navigator 520 (0.42 ± 0.17; p-value = 0.03) and the Pixel 6 (0.36 ± 0.18; p-value < 0.01). For performance (Question 4), the Trimble XR10 did not significantly improve performance quality (0.64 ± 0.20) compared to the Navigator 520 (0.43 ± 0.19; p-value < 0.01) and the Pixel 6 (0.41 ± 0.19; p-value < 0.01). In terms of satisfaction (Question 5), Trimble XR10 users reported lower performance satisfaction (0.57 ± 0.20) compared to the Navigator 520 (0.41 ± 0.18; p-value = 0.03) and the Pixel 6 (0.40 ± 0.20; p-value = 0.02).

Regarding safety and well-being (Question 6), Trimble XR10 users felt the device did not improve their safety and well-being (0.68 ± 0.16) compared to the Navigator 520 (0.43 ± 0.16; p-value < 0.01) and the Pixel 6 (0.49 ± 0.22; p-value < 0.01). Regarding discomfort (Question 8), Trimble XR10 users reported more discomfort (0.59 ± 0.20) compared to Navigator 520 users (0.79 ± 0.23; p-value < 0.01) and Pixel 6 users (0.85 ± 0.17; p-value < 0.01). For learning (Question 7) and workflow (Question 9), Navigator 520 users felt it significantly promoted learning and skill expansion (0.36 ± 0.14) compared to the Pixel 6 (0.56 ± 0.23; p-value < 0.01). Conversely, Trimble XR10 users perceived less improvement in workflow (0.61 ± 0.15) compared to the Navigator 520 (0.43 ± 0.16; p-value = 0.01).

\begin{figure*} [tb]
\centering
\includegraphics[width=0.6\linewidth]{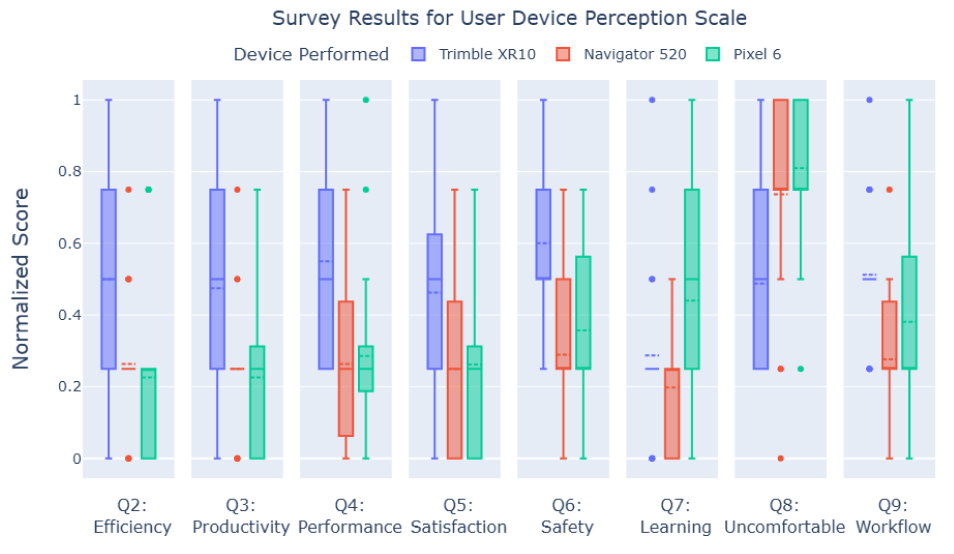}
    \centering
    \captionof{figure}{Normalized User Device Perception scores reveal statistically significant differences between conditions. The Trimble XR10 showed lower scores in efficiency, productivity, performance, satisfaction, safety, and comfort, while the Navigator 520 facilitated better learning.}
    \label{fig:6}
\end{figure*}

Among the three conditions evaluated (see Figure \ref{fig:6}), the Trimble XR10 received the poorest device perception scores in terms of productivity, performance, satisfaction, safety, well-being, and comfort enhancement compared to both Navigator 520 and Pixel 6 conditions. Conversely, there were no significant differences between Navigator 520 and Pixel 6 across these aspects. However, Navigator 520 showed significant differences in promoting learning and skill expansion compared to Pixel 6 (\textbf{RQ3}).

These results differ from other works that have reported improvements in subjective performance, productivity, and efficiency while using the Trimble XR10 for industrial tasks \cite{REVOLTI2023746}. This discrepancy could be attributed to differences in experimental conditions. Many previous studies have focused on static industrial workstations. In contrast, our study evaluated a more complex environment that included navigation tasks. It is well-established that navigation tasks are affected by the cognitive load \cite{Succesfulnavigation}. Additionally, incorporating hazardous elements that require user awareness \cite{PinCognitive} could further impact performance, satisfaction, and the efficiency perceptions of the device. A similar study in a less complex industrial environment found that the Trimble XR10 did not expedite task completion or enhance job satisfaction. \cite{Howard2023VisualInspectionWithAugmentedRealityHMD}. Future studies should investigate the effects of off-the-shelf AR-HMDs in high-cognitive-demand environments to gain insights into design approaches that can mitigate the inherent complexities of AR interface interactions (e.g., unnatural selection methods such as pinching the air) in settings where cognitive load demands are already high.

\subsection{System Usability Scale - SUS}

In this study, the SUS was employed to measure participants' subjective perceptions of the system's usability. The scale consists of ten Likert scale (1—Strongly Disagree to 5—Strongly Agree) that capture users' impressions of the system's usability, effectiveness, efficiency, and learnability. Table~\ref{tab:SubjectiveSummaries} presents the summarized results for each SUS question.

\begin{figure*} [tb]
\centering
\includegraphics[width=1\linewidth]{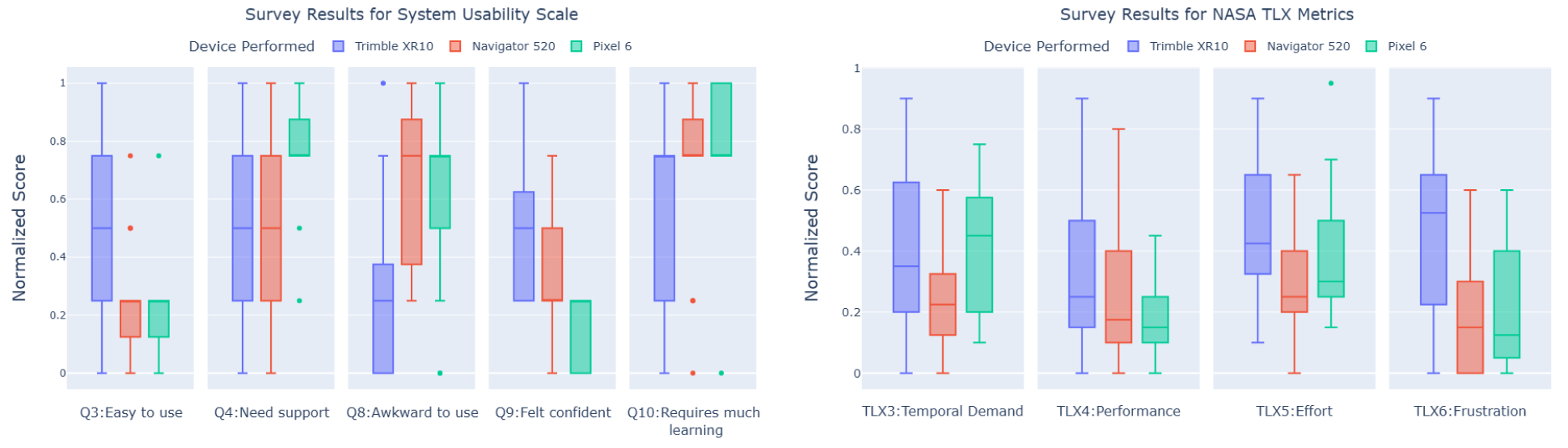}
    \centering
    \captionof{figure}{Normalized System Usability Scale and NASA TLX scores that reported statistically significant differences between the conditions.}
    \label{fig:7}
\end{figure*}

Figure \ref{fig:7} presents the SUS results that showcased stastisticaly significant differences between the conditions. Regarding the ease of use (Question 3), participants using the Trimble XR10 found the interface significantly more challenging to use (0.59 ± 0.23) compared to those using the Navigator 520 (0.40 ± 0.16; p-value < 0.01) and Pixel 6 (0.37 ± 0.13; p-value < 0.01). For technical support needed to use the device (Question 4), Trimble XR10 users reported requiring significantly more assistance (0.62 ± 0.26) compared to Pixel 6 users (0.82 ± 0.14; p-value = 0.01). Similarly, Navigator 520 users also reported needing more assistance (0.61 ± 0.24) compared to Pixel 6 users (0.82 ± 0.14; p-value < 0.01). Additionally, for the amount of learning required to use the device (Question 10), Trimble XR10 users reported needing to learn significantly more (0.68 ± 0.23) compared to Pixel 6 users (0.85 ± 0.18; p-value = 0.04). Regarding awkwardness (Question 8), users found the Trimble XR10 interface notably more awkward (0.44 ± 0.24) than the Navigator 520 (0.72 ± 0.23; p-value < 0.01) and the Pixel 6 condition (0.71 ± 0.25; p-value < 0.01). Regarding confidence (Question 9), users in the Trimble XR10 reported significantly less confidence while using the interface (0.58 ± 0.18) compared to those in the Navigator 520 (0.43 ± 0.16; p-value < 0.01) and Pixel 6 (0.33 ± 0.1; p-value < 0.01). 

Our results indicate significant differences in usability between the devices, with the Trimble XR10 being the most challenging and awkward to use among the conditions (\textbf{RQ3}). Participants in our study reported needing more technical assistance with this device and feeling less confident while using it. This contrasts with the more positive findings reported by \cite{Howard2023VisualInspectionWithAugmentedRealityHMD}. The complexity of the tasks and environments in our study, particularly the presence of multiple hazards, may explain these differing outcomes. Observational notes from the training sessions suggest that the interaction methods of the Navigator 520 (i.e., voice commands) and the Pixel 6 (i.e., touch interaction) were more intuitive for users. In contrast, the Trimble XR10's mid-air interaction felt unnatural to participants. This interaction method presented a steeper learning curve and more technical difficulties, which ultimately impacted the overall user experience negatively. This is consistent with the findings from other works on mid-air interactions \cite{NeateMidAir}. For instance, participants noted that the lack of haptic feedback affected their perception of distance and object selection compared to physical manipulation. Furthermore, similar to the findings by \cite{Egger2020AugmentedRealityinSupportofIntelligentManufacturing}, participants experienced tracking issues, where the system struggled to accurately detect hand gestures, especially while wearing protective gloves as part of their PPE.

Future work should aim to create a more natural interaction experience, such as incorporating haptic feedback to simulate physical manipulation when interacting with panels. This could enhance the overall user experience. Additionally, addressing hardware limitations related to hand tracking while wearing gloves is essential. 

\subsection{Task Load Index - TLX}

The NASA Task Load Index (NASA TLX) was employed to assess participants' subjective perceptions of the workload experienced during the tasks. This index comprises six scales that evaluate different aspects of workload: Mental Demand, Physical Demand, Temporal Demand, Performance, Effort, and Frustration. Participants rate each item on a scale from 0 (very low) to 21 (very high), providing insights into the cognitive, physical, and emotional demands experienced while performing the assigned tasks. Table~\ref{tab:SubjectiveSummaries} presents the summarized results for each TLX question.

For temporal demand (TLX3), Trimble XR10 users felt significantly more rushed (0.43 ± 0.25) compared to Navigator 520 users (0.27 ± 0.15; p-value = 0.04). Additionally, regarding performance (TLX4), Trimble XR10 users felt significantly less successful in accomplishing their tasks (0.38 ± 0.25) compared to Pixel 6 users (0.21 ± 0.11; p-value = 0.03). Regarding effort (TLX5), users of the Trimble XR10 reported expending significantly more effort to achieve their level of performance (0.49 ± 0.93) compared to those using the Navigator 520 (0.31 ± 0.18; p-value = 0.01). Additionally, concerning frustration (TLX6), Trimble XR10 participants reported feeling significantly more frustrated (0.48 ± 0.25) than those using the Navigator 520 (0.22 ± 0.17; p-value < 0.01) and the Pixel 6 (0.26 ± 0.20; p-value < 0.01).

Overall, our results show significant differences between the Trimble XR10, the Navigator 520, and the Pixel 6 (see Figure \ref{fig:7}). The Trimble XR10 demonstrated statistically lower performance and higher levels of temporal demand, effort, and frustration during industrial tasks (\textbf{RQ3}). These findings contrast with those reported by \cite{Blattgerste2017ComparingConventionalAndAugmentedRealityInstructionsAssembly, Hou2013UsinngAnimatedAugmentedRealityGuideAssembly}, where AR devices led to shorter task completion times, fewer assembly errors, and lower overall task load in a stationary LEGO assembly task. Moreover, the phone condition in those studies was associated with higher cognitive load scores, which contradicts our findings. 

The discrepancy may be explained by the presence of integrated hazards and acute stressors in our study, which have been shown in other research to affect cognitive load, impair memory, and cause attention narrowing \cite{Gloria2020}. These factors, along with the learning curve associated with AR devices, likely contributed to the differences observed in the TLX scores. Our findings emphasize the need to evaluate AR technologies in more realistic and dynamic industrial environments. Designers should consider the cognitive load imposed by the complexity of these settings when developing interfaces and explore strategies to reduce this load to enhance user safety.

\subsection{User Safety Perception - USP}

A Likert scale questionnaire (1—Strongly Agree to 5—Strongly Disagree) was usnoeed to evaluate the subjective safety perception of the devices. The results of this User Safety Perception test, particularly Question 6, indicate that Trimble XR10 users reported feeling significantly more uneasy or insecure while navigating the room (0.74 ± 0.22) compared to Navigator 520 users (0.90 ± 0.10; p = 0.04). Table~\ref{tab:SubjectiveSummaries} presents the summarized USP results.

This suggests that the Trimble XR10 was perceived as less safe compared to the Navigator 520 (see Figure \ref{fig:6}), which refers back to (\textbf{RQ3}). As mentioned earlier, the weight difference between devices and their intrusiveness may impact the sense of discomfort \cite{RichardMuscleForce}. This sense of discomfort can increase cognitive load, which in turn affects the user's memory, situational awareness, and attentional focus. Since these cognitive factors are already influenced by the complexity of the environment, heightened discomfort may further reduce the perception of safety \cite{Constantino2021NewAnEmergingHazardsForHealthAndSafetyWithinDigitalizedManufacturing,Wei2023HeadMountedDisplayAugmentedRealityInManufacturing}.

Moreover, participants expressed discomfort regarding the constant display of the interface on the Trimble XR10, in contrast to the Navigator 520 where participants could choose when to view the interface. Participants reported that the persistent visibility of the interface occasionally obstructed their field of view while walking. As one participant described, \textit{"The fact that the panel required high maintenance sometimes drew my attention away from my surroundings. At times, it blocked my vision when it did not track well"}.

Further studies should address the ergonomic impacts of different AR-HMD devices (e.g., muscular tension, and adoption of unnatural positions). Likewise, manufacturers of AR-HMD devices should focus on reducing device weight without compromising their capabilities. Concerning UI/UX, designers need to be cautious with the placement of interfaces, especially in hazardous settings. We suggest that designers delve deeper into the development of adaptive interfaces. These interfaces should be capable of analyzing environmental context and making real-time design decisions to reduce users' cognitive load and minimize occlusion. By doing so, they can enhance safety in industrial settings. 

\subsection{Hazard Identification}

For the Hazard Identification Questionnaire, we asked each participant to identify any hazards they recalled encountering during their run-through of the experiment. Nonexistent hazards, e.g., from falling from height to moving hazards, were included to account for participants reporting ``yes" to all hazards. As seen in Table \ref{tab:hazards}, each column represents the condition, and each row represents a hazard type. Based on the results, no statistically significant differences were found between the conditions. The study findings indicate that the type of AR-HMD device does not significantly impact the users' ability to recognize hazards in an industrial setting (\textbf{RQ2}).

\begin{table}[ht]
\centering
\caption{Total Number of Identified Hazards per Condition}
\begin{tabular}{|l|l|l|l|}
\hline
\textbf{Type of Hazard}      & \textbf{Trimble} & \textbf{Navigator} & \textbf{Pixel} \\ \hline
\hline
Pipe Hazards        & 18           & 18            & 18      \\ \hline
Head Knockers       & 17           & 17            & 19      \\ \hline
Slipper Surfaces    & 15           & 18            & 18      \\ \hline
Noise Hazards       & 15           & 16            & 18      \\ \hline
Holes and Gaps      & 12           & 16            & 16      \\ \hline
Uneven Surfaces     & 19           & 18            & 16      \\ \hline
Knee Knockers       & 12           & 16            & 13      \\ \hline
Confined Spaces     & 12           & 19            & 15      \\ \hline
Obstruction Hazards & 12           & 14            & 12       \\ \hline
Obstructed Vision   & 11           & 9             & 8       \\ \hline
High Temperatures   & 5            & 2             & 3       \\ \hline
\hline
Falling from Height & 5            & 4             & 2       \\ \hline
Falling Objects     & 2            & 1             & 4       \\ \hline
Chemical Exposures  & 2            & 9             & 4       \\ \hline
Fire and Explosion  & 3            & 4             & 3       \\ \hline
Moving Hazards      & 0            & 1             & 1       \\ \hline
\hline
Total               & 160          & 182           & 170       \\ \hline

\end{tabular}
\label{tab:hazards}
\end{table}

\subsection{Recommendations for AR-HMDs within Industrial Hazardous Environments}
The results show that although wearing an AR-HMD did not show significant differences in the users' ability to recognize hazards in an environment (\textbf{RQ2}), wearing the devices did significantly lead to more sensor activations, i.e., potential accidents, particularly hazards related to head and knee knockers (\textbf{RQ1}). Moreover, the Trimble XR10 had poorer scores in productivity, performance, satisfaction, safety, well-being, and comfort, along with higher temporal demand, effort, frustration, and greater difficulty of use (\textbf{RQ3}). These results confirm that current AR-HMD technology is not ready to adequately ensure the SSA of their users within industrial hazardous environments without compromising the situation awareness.

In light of these findings, the limitations of off-the-shelf AR-HMDs within industrial hazardous environments were quantified. The results point towards their negative impacts on the workers’ safety and ability to perceive the environment properly. Although our analysis was focused on specific AR devices, we believe these are representative of the bigger spectrum of commercially-available devices, i.e., respectively, the Trimble XR10, the Navigator 520, and the Pixel 6 are representative examples of stereo see-through displays, multiplexed monocular displays, and mobile handheld devices. As such, our intention is to generalize the findings from this work to the bigger spectrum of AR devices.

Despite the stereo see-through display showing overall lower results compared to the monocular HMD and handheld devices, industries are rapidly adopting this type of devices thanks to their proven performance improvements \cite{Wei2023HeadMountedDisplayAugmentedRealityInManufacturing}. In contrast, while handheld devices are more familiar and easier to implement, the drive to adopt them in industrial environments is decreasing: keeping workers' hands occupied can impair their reactions and situational awareness, potentially leading to hazardous situations over time \cite{DANIELSSONOperatorsperspective, SYBERFELDTShopfloor}.

Given these adoption trends and limitations, future research should focus on incorporating safety features into stereo see-through AR-HMDs. These algorithms should combine Computer Vision, Artificial Intelligence, and Computer Graphics to maximize the efficiency of AR technologies in industrial settings. The objective should be to develop AR enhancements that can sustain and reinforce the workers’ safety and situational awareness as they perform their tasks. Three possible advancements could be developed. First, the AR-HMDs' understanding of the users should be addressed. AR-HMDs should assess the user’s current level of stress and distraction while performing a task. If these levels are too high, the system should assist the user to bring those levels back to normal range. Second, AR-HMDs should passively understand the environment and adapt the UI accordingly, e.g., changing the color and contrast with the background, size relative to the users’ position, and placement relative to the user's view. Through these changes, users could spend more time focused on the tasks instead of dealing with the UI's configuration within their FOV. Finally, the AR-HMDs' active understanding of the environment should be improved to recognize potential hazards and notify the user of any possible dangers without introducing additional risks to the user.

\section{Conclusion}

This paper presented a comprehensive analysis of whether current AR-HMDs adequately ensure safety within industrial hazardous environments without compromising the situation awareness of the workers. Our work addressed the lack of comprehensive analyses, i.e., quantitative and qualitative, showcasing how much the workers' safety and situation awareness gets reduced while wearing AR-HMDs. Our evaluation was comprised of sixty participants performing various tasks in a simulated hazardous environment while receiving remote expert guidance through one of three devices: two off-the-shelf AR-HMDs and a traditional smartphone approach. 
The results pointed out that the Trimble XR10 led to statistically higher potential safety incidents compared to the Navigator 520 and the Pixel 6. The results revealed that wearing the AR-HMDs led to a significantly perceived lower ability to recognize hazards, perceived safety, comfort, perceived performance, and usability. Overall, these results indicate that AR-HMDs need to be enhanced further before they can be successfully integrated into industrial hazardous settings without compromising the workers' safety.

\section{Acknowledgements}
This work was supported under a Texas A\&M University - American Bureau of Shipping research agreement and the Laboratory of Ocean Innovation at Texas A\&M University. Opinions, interpretations, conclusions and recommendations are those of the author and are not necessarily endorsed by the funders.

\begin{table}[ht]
\centering
\caption{Summary of normalized answers to subjective questionnaires. Results are reported as mean ± standard deviation. Metrics reporting statistical significant differences are highlighted in gray}
\begin{tabular}{|l|l|l|l|}
\hline
\textbf{Metrics}      & \textbf{Trimble XR10} & \textbf{Navigator 520} & \textbf{Pixel 6} \\ \hline
\hline
UDP1 & 0.42 ± 0.14 & 0.35 ± 0.13 & 0.32 ± 0.16 \\ \hline
\rowcolor{LightGray}
UDP2 & 0.60 ± 0.27 & 0.43 ± 0.18 & 0.36 ± 0.18 \\ \hline
\rowcolor{LightGray}
UDP3 & 0.58 ± 0.23 & 0.42 ± 0.17 & 0.36 ± 0.18 \\ \hline
\rowcolor{LightGray}
UDP4 & 0.64 ± 0.20 & 0.43 ± 0.19 & 0.41 ± 0.19 \\ \hline
\rowcolor{LightGray}
UDP5 & 0.57 ± 0.20 & 0.41 ± 0.18 & 0.40 ± 0.21 \\ \hline
\rowcolor{LightGray}
UDP6 & 0.68 ± 0.16 & 0.43 ± 0.16 & 0.49 ± 0.22 \\ \hline
\rowcolor{LightGray}
UDP7 & 0.43 ± 0.20 & 0.36 ± 0.14 & 0.56 ± 0.23 \\ \hline
\rowcolor{LightGray}
UDP8 & 0.59 ± 0.20 & 0.79 ± 0.23 & 0.85 ± 0.17 \\ \hline
\rowcolor{LightGray}
UDP9 & 0.61 ± 0.15 & 0.43 ± 0.16 & 0.50 ± 0.24 \\ \hline
\hline
SUS1 & 0.57 ± 0.24 & 0.45 ± 0.17 & 0.45 ± 0.19 \\ \hline
SUS2 & 0.76 ± 0.18 & 0.82 ± 0.16 & 0.83 ± 0.15 \\ \hline
\rowcolor{LightGray}
SUS3 & 0.59 ± 0.23 & 0.40 ± 0.16 & 0.37 ± 0.13 \\ \hline
\rowcolor{LightGray}
SUS4 & 0.62 ± 0.26 & 0.61 ± 0.24 & 0.82 ± 0.14 \\ \hline
SUS5 & 0.46 ± 0.18 & 0.38 ± 0.17 & 0.51 ± 0.20 \\ \hline
SUS6 & 0.67 ± 0.25 & 0.78 ± 0.20 & 0.80 ± 0.13 \\ \hline
SUS7 & 0.43 ± 0.13 & 0.39 ± 0.19 & 0.38 ± 0.18 \\ \hline
\rowcolor{LightGray}
SUS8 & 0.44 ± 0.24 & 0.72 ± 0.23 & 0.71 ± 0.25 \\ \hline
\rowcolor{LightGray}
SUS9 & 0.58 ± 0.18 & 0.43 ± 0.16 & 0.33 ± 0.10 \\ \hline
\rowcolor{LightGray}
SUS10 & 0.68 ± 0.23 & 0.76 ± 0.23 & 0.85 ± 0.18 \\ \hline
\hline
TLX1 & 0.45 ± 0.23 & 0.36 ± 0.16 & 0.42 ± 0.18 \\ \hline
TLX2 & 0.35 ± 0.37 & 0.27 ± 0.16 & 0.36 ± 0.20 \\ \hline
\rowcolor{LightGray}
TLX3 & 0.43 ± 0.25 & 0.27 ± 0.15 & 0.43 ± 0.20 \\ \hline
\rowcolor{LightGray}
TLX4 & 0.38 ± 0.25 & 0.29 ± 0.21 & 0.21 ± 0.11 \\ \hline
\rowcolor{LightGray}
TLX5 & 0.49 ± 0.93 & 0.31 ± 0.18 & 0.41 ± 0.19 \\ \hline
\rowcolor{LightGray}
TLX6 & 0.48 ± 0.25 & 0.22 ± 0.17 & 0.26 ± 0.20 \\ \hline
\hline
USP1 & 0.40 ± 0.19 & 0.35 ± 0.19 & 0.32 ± 0.12 \\ \hline
USP2 & 0.78 ± 0.18 & 0.82 ± 0.15 & 0.69 ± 0.25 \\ \hline
USP3 & 0.72 ± 0.23 & 0.85 ± 0.16 & 0.79 ± 0.19 \\ \hline
USP4 & 0.41 ± 0.18 & 0.38 ± 0.18 & 0.36 ± 0.17 \\ \hline
USP5 & 0.41 ± 0.21 & 0.37 ± 0.19 & 0.38 ± 0.19 \\ \hline
\rowcolor{LightGray}
USP6 & 0.74 ± 0.22 & 0.90 ± 0.10 & 0.77 ± 0.25 \\ \hline
USP7 & 0.91 ± 0.12 & 0.88 ± 0.10 & 0.87 ± 0.16 \\ \hline
USP8 & 0.53 ± 0.25 & 0.44 ± 0.25 & 0.47 ± 0.23 \\ \hline
USP9 & 0.46 ± 0.17 & 0.39 ± 0.18 & 0.43 ± 0.19 \\ \hline
USP10 & 0.80 ± 0.16 & 0.83 ± 0.12 & 0.84 ± 0.12 \\ \hline

\end{tabular}
\label{tab:SubjectiveSummaries}
\end{table}


\bibliographystyle{abbrv-doi-hyperref}

\bibliography{template}

\end{document}